# Two-dimensional coupled photonic crystal resonator arrays

Hatice Altug[*] and Jelena Vučković[†]

*Edward L. Ginzton Laboratory, Stanford University, Stanford, CA 94305-4088*

**Abstract**

We present the design and fabrication of photonic crystal structures exhibiting electromagnetic bands that are flattened in all crystal directions, i.e., whose frequency variation with wavevector is minimized. Such bands can be used to reduce group velocity of light propagating in arbitrary crystal direction, which is of importance for construction of miniaturized tunable optical delay components, low-threshold photonic crystal lasers, and study of nonlinear optics phenomena.

[*] Also at the Department of Applied Physics, Stanford University, Stanford, CA 94305, altug@stanford.edu
[†] Also at the department of Electrical Engineering, Stanford University, Stanford, CA 94305, jela@stanford.edu, http://ee.stanford.edu/~jela/jv_files/group.html

Small group velocity is crucial in variety of applications, ranging from optical delay components and low-threshold lasers, to the study of nonlinear optics phenomena. Although photonic crystals [1,2] can be employed to achieve slow group velocities at their band edges [3,4], this is limited to a very narrow range of wavevectors in one particular direction. Moreover, a large variation of the group velocity with wavevector (group velocity dispersion) implies a significant distortion in the shape of an optical pulse propagating through the structure [5]. Recently, coupled resonator optical waveguides (CROW's) in photonic crystals were also proposed for reducing the group velocity in a wide range of wavevectors, but still only for propagation in a narrow region along the waveguide axis [6].

In this letter, we describe design and fabrication of two-dimensional (2D) arrays of coupled photonic crystal resonators that exhibit flat bands (reduced group velocity) over the entire range of wavevectors and in all crystal directions. This decreases the sensitivity of coupling and minimizes distortion of an optical pulse propagating through the structure. Beside construction of optical delay components, this proposal is also of importance for building low threshold photonic crystal lasers with increased output powers. Demonstrated distributed feedback photonic crystal lasers [7] can provide larger output powers than photonic crystal micro-cavity lasers, as they rely on high density of optical states near the band edge. However, it has been predicted theoretically that the reduction in the group velocity over a wide range of wavevectors (so called group velocity anomaly) would lead to an even larger decrease in the lasing threshold [8], and this has been in part employed in lasing from one-dimensional (1D) CROW structures [9]. As a result of a large group velocity anomaly in any photonic crystal direction, our proposal can enable even higher output powers while preserving low lasing thresholds. Likewise, electrical pumping of these structures is not as formidable task as in presently pursued photonic crystal microcavity lasers. Finally, in combination with nonlinear optical materials, the proposed 2D coupled photonic crystal resonator arrays (CPCRA) could also be used for exploration of 2D discrete solitons [10,11] or construction of optical switching arrays.

2D CPCRA's studied in this article are constructed by periodically modifying holes of a square lattice photonic crystal slab; for example, every third lattice hole in both $x$ and $y$ directions can be removed, as shown in Fig. 1a. Fig. 1b shows the unit cell and directions of high-symmetry points, Γ, X and M of the CPCRA from Fig. 1a. This structure can be viewed as a 2D array of single-defect photonic crystal cavities formed by removing a single air hole. An isolated cavity of this type supports three types of modes: doubly degenerate dipole, non-degenerate quadrupole and non-degenerate monopole. As the lattice perturbation increases (i.e., the defect hole radius decreases) modes are pulled from the air band and localized in the band gap. The first mode to be localized in this process is the dipole, and the last is the monopole. The cavity mode with the highest quality factor (Q-factor) is the quadrupole that has been discussed in more detail in Ref. 12.

All theoretical results presented in this letter were obtained by the three-dimensional finite-difference time-domain (3D FDTD) method [13], and the spatial resolution in discretization of the structure is 20 computational points per interhole spacing $a$. The thickness of the slab is $d=0.55a$, the hole radius is $r=0.4a$, and the refractive index is assumed to be $n=3.5$, corresponding to silicon (Si) at optical wavelengths. The application of Bloch boundary conditions (BC) to four edges in the $x$- and $y$-directions allows the analysis of only one unit cell of the structure, and the use of the mirror BC at the lower boundary in the $z$-direction (in the middle of the slab) reduces the computation time and enables the excitation of Transverse electric (TE) field-like (even) modes only. The frequency resolution of our calculations is $2.2 \cdot 10^{-3}$ in normalized units of $a/\lambda$, where $\lambda$ corresponds to the mode wavelength in air.

When cavities are tiled together in the CPCRA, coupled cavity bands form. Band diagram of the structure from Fig. 1a is shown in Fig. 2. Since its unit cell (shown both in Fig. 1b) is three times bigger than the unit cell of the original square lattice, it is necessary to fold the square lattice band diagram three times in order to plot it together with the band diagram of the CPCRA. In Fig. 2 we assume that positions and shapes of the guided bands of the original square lattice photonic crystal remain unchanged inside the CPCRA. This is only an approximation, as their frequencies may decrease due to the increased overlap with high-index material, and new minigaps may open at high symmetry points of the CPCRA. Dielectric and air bands are shaded in gray, as we are primarily interested in the modes located in the band gap; these, in turn, are the portions of the band diagram that are easily accessible experimentally. Although the coupled dipole bands are located below the top air band edge in the XM direction, they are included in the diagram to clarify the explanation of the coupled mode characteristics. The dipole mode is doubly degenerate in an isolated cavity [12,13], but it splits into two bands corresponding to the $x$- and $y$-dipoles in the CPCRA structure, as shown in Fig. 2. The $x$ ($y$) dipole has electric field polarized in the $x$ ($y$) direction at the center of the defect and radiates mostly in the *y* (*x*) direction. The inset (d) of Fig. 2 shows the *z*-component, the only non-zero component of magnetic field in the middle of the slab, for the *x*-dipole band at X point. Due to the previously described radiation pattern, the *x*-dipole is coupled more in the XM direction than in the ΓX direction, and its band is thus flatter in the ΓX than in the XM direction. The situation is opposite for the *y*-dipole. The preferential coupling of the dipole mode in one lattice direction implies that its coupled band properties are similar to a 1D CROW case. In the ΓM direction, *x*- and *y*- dipoles combine and form a diagonal dipole. We have also observed a coupled monopole band in this CPCRA structure, whose magnetic field pattern at the Γ point is shown in the inset (c) of Fig. 2. Due to its poor confinement and Q-factor, there is a significant slope in the coupled band.

The coupled band of our primary interest is the one corresponding to the quadrupole mode. The insets (a) and (b) of Fig. 2 show magnetic field patterns of this mode at X and M points, respectively. Quadrupole radiates equally in the four ΓM directions and its radiation pattern thus has a four-fold-symmetry; this also implies that the mode couples equally to all of its four neighbors in a particular lattice direction (e.g., ΓM or ΓX). The lack of preferential coupling directions and a good lateral confinement (high Q-factor) lead to a flat coupled-quadrupole band. In addition, this mode is non-degenerate and does not split into several sub-bands as the coupled-dipole band.

The coupled-quadrupole band can be made even "flatter" by inserting a larger number of photonic crystal periods between the defects, and thereby reducing the coupling between individual resonators, as proposed in the 1D CROW case [6]. Fig. 3 shows the coupled-quadrupole band for 2D CPCRA structure with three and four-rows of photonic crystal between cavities, while all other photonic crystal parameters (*d*, *a*, and *r*) are the same as previously. As the number of photonic crystal rows between cavities increases from two to four, the band becomes flatter, and the frequency difference between its maximum and minimum is reduced from $6.6 \cdot 10^{-3}$ to $2.2 \cdot 10^{-3}$ (the frequency resolution of our FDTD simulation). Moreover, it is possible to control the position of this band within the band gap by employing defects consisting of holes with reduced radius. This increases the overlap of the field with air region inside an individual resonator, and leads to an increase in the mode frequency. As an illustration of this effect, Fig. 3 shows a coupled- quadrupole band of the same three-row CPCRA, but with reduced effect air holes radius ($r_d$=0.3$a$) inserted, whose frequency is significantly higher relative to the three-row CPCRA with completely removed defect-holes ($r_d$=0). Coupled-quadrupole bands for even and odd number of photonic crystal rows between cavities are convex and concave, respectively; this is a result of the difference in refractive index at midpoints between resonators, i.e., different overlap of the coupled mode field with low and high-index regions.

Finally, in order to demonstrate the feasibility of our proposal, we have fabricated the designed CPCRA structure in silicon on insulator (SOI) for an operating wavelength of 1550 nm. The desired thickness of the Si membrane can be achieved by wafer thermal oxidization followed by the HF-wet etching of the formed oxide layer. Fabrication process starts with spinning of the PMMA layer with molecular weight of 495K on the sample surface, followed by baking on a hot plate at 170$^o$C for 30 min; this results in the measured PMMA thickness of 320 nm. Electron-beam lithography is then performed in Raith 150 system at 10 keV, and the exposed PMMA is developed in 3:1 IPA:IMBK mixture for 50 seconds and rinsed in IPA for 30 sec. Patterns are subsequently transferred to Si using magnetically induced reactive ion etch with HBr/Cl$_2$ gas combination. After dry etching, remaining PMMA is removed by O$_2$ plasma process. Finally the oxide layer underneath Si was removed by immersing the sample into the buffered HF, which leads to a freestanding Si membrane. Fabricated structures are uniform over large area, as can be confirmed in the SEM pictures shown in Fig. 1c. The experimental study of the fabricated structures is in progress and will be reported in our forthcoming publication.

In conclusion, we have shown that 2D CPCRA can be used to achieve bands exhibiting slow group velocity over all wavevectors and in all crystal directions. Design can be modified to tune the band slope and its frequency within the band gap. This proposal is important for construction of low-threshold photonic crystal lasers with increased output power, optical delay components, and study of nonlinear optics phenomena. Furthermore, use of refractive index modulation in such structures could enable dynamically tunable group velocity and realization of optical devices that could store and release optical pulses. The study presented here can be extended to three dimensional coupled resonator arrays, as well as to other types of resonators, including those not based on photonic crystals.

**Acknowledgement** This work has been supported in part by the MURI Center for Photonic Quantum Information Systems (ARO/ARDA program DAAD19-03-1-0199), the Charles Lee Powell Foundation Faculty Award and the Terman Award.

**Figure Captions**

**Figure 1.** (a) Schematic configuration of the simulated and fabricated structure. (b) Unit cell and location of high symmetry points. (c) SEM picture of the fabricated two-layer CPCRA.

**Figure 2.** Band diagram for two layer 2D CPCRA structure (shown in Fig 1.) for TE-like modes only. Dielectric and air bands of the original square lattice are folded three times separately in ΓX, XM and Γ M and shaded in gray. Magnetic field pattern along the *z*-direction for: (a) quadrupole mode at X point; (b) quadrupole mode at M point; (c) monopole mode at Γ point; (d) *x*-dipole mode at X point. Unit cell in real space and location of high symmetry points in the reciprocal space are also shown in the inset.

**Figure 3.** Band diagram for the coupled quadrupole band in the CPCRA with two, three, or four photonic crystal rows between the resonators, and with resonators consisting of either completely removed single holes ($r_d/a = 0$) or with reduced radius holes ($r_d/a = 0.3$). Insets show unit cells of three and four row structures in real space, and locations of high symmetry points in the reciprocal space. Regions below the top edge of the dielectric band and above the bottom edge of the air band of the original square lattice are shaded in gray.

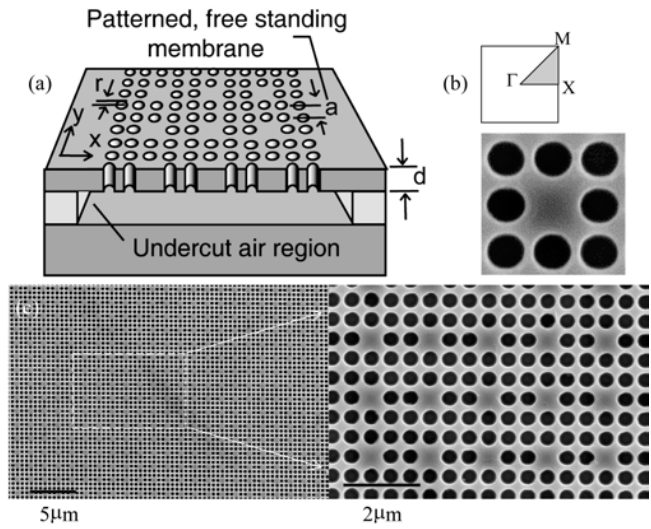

**FIGURE 1**

**Authors: H. Altug, J. Vučković**

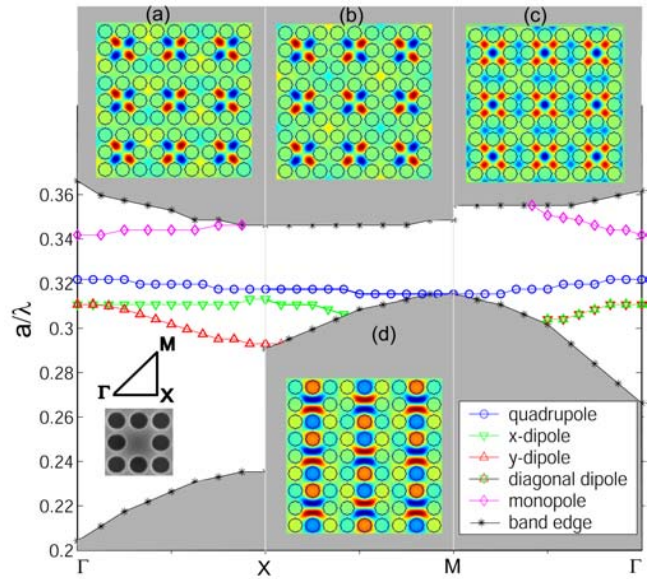

**FIGURE 2**

**Authors: H. Altug, J. Vučković**

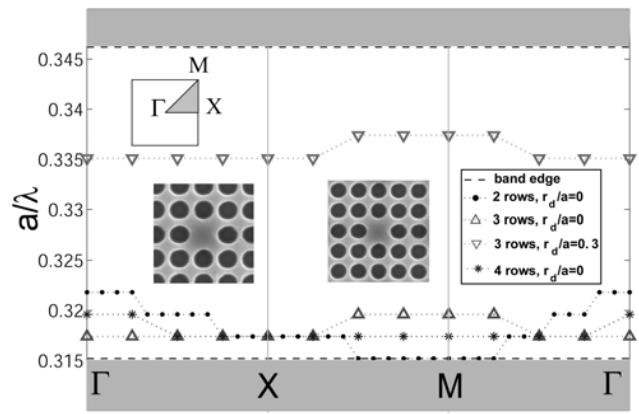

**FIGURE 3**

**Authors: H. Altug, J. Vučković**